\documentstyle[epsbox]{res}
\begin{document}

\title{%
Dark Matter Halos Simulated with Million Particles}

\author{Y.P. JING \\%% <== First author
{\it National Astronomical Observatory, Mitaka,
 Tokyo 181-8588, Japan\\ jing@utap.phys.s.u-tokyo.ac.jp}\\
%{\it Research Center for the Early Universe, University of
%       Tokyo\\
% Bunkyo-ku, Tokyo 113-0033, Japan, jing@utap.phys.s.u-tokyo.ac.jp}\\
Yasushi SUTO \\%% <== Second author
{\it Research Center for the Early Universe, University of
       Tokyo\\
 Bunkyo-ku, Tokyo 113-0033, Japan, suto@phys.s.u-tokyo.ac.jp}}

\maketitle

\section*{Abstract}

We report a series of high-resolution N-body simulations designed to
examine the internal physical properties of dark matter halos. A total
of fifteen halos, each represented by $\sim 1$ million particles
within the virial radius, have been simulated covering the mass range
of $2\times10^{12}\sim 5\times10^{14} h^{-1}{\rm M_\odot}$.  As the
first application of these simulations, we have examined the density
profiles of the halos. We found a clear systematic correlation between
the halo mass and the slope of the density profile at one percent of
the virial radius, in addition to the variations of the slope among
halos of the similar mass. More specifically, the slope is $\sim -
1.5$, $-1.3$, and $-1.1$ for galaxy, group, and cluster mass halos,
respectively. Thus we conclude that the dark matter density profiles,
especially in the inner region, are not universal.

\section{Introduction}

The internal structures within dark matter halos have important impact
on galaxies and clusters formed within these halos. This is the reason
for recent enormous interest in the innermost density profiles of
the dark matter halos. From a set of halo simulations with $10^4$
particles, Navarro, Frenk \& White (1997; NFW) concluded that the
density profiles universally obey the NFW form $\rho(r)\propto
r^{-1}(r+r_s)^{-2}$. It is yet unclear to which degree their results
are affected by their selection criterion which is not well-defined,
as questioned recently by Jing (1999). With a higher resolution of
$\sim 10^6$ particles, Fukushige \& Makino (1997) and Moore et
al. (1998) found a steeper inner slope and argued that the slope might
converge to a value about $-1.5$. However, Kravtsov et al.(1998), with
a resolution in between the above studies as the authors themselves
claimed, found a shallower slope than the NFW one. Therefore, a
consensus still needs to be reached about the inner density profile of
the halos.

Motivated by these confusing results and the importance of
studying other internal properties of the halos, e.g. substructures,
we have selected 15 halos from a cosmological simulation and
resimulated each halo with $\sim 10^6$ particles. We emphasize our
selection of the halos is random (cf. NFW), in order to pin down
selection effects introduced to the sample, thus in order to unbiasly
study the internal properties of the dark matter halos.
 
\section{Simulation procedure}

We adopt the following two-step procedure. First we select 15 dark
matter halos from our LCDM cosmological N-body simulation (Jing \&
Suto 1998), with five in each category which has a mass scale of
clusters, groups, and galaxies respectively. These halos are
resimulated typically with $2.2\times 10^6$ particles, $\sim 1.5\times
10^6$ fine particles for the central halo region and its nearby
outskirts, and $\sim 0.7\times 10^6$ coarse particles for outer region.
About $(0.5 - 1)\times 10^6$ particles end up within the virial radius
of each halo, making this sample as the largest sample of million
particle halo simulations.

We developed a new algorithm, the nested-grid ${\rm P^3M}$ algorithm,
to calculate the force with an accuracy around 0.1 percent. The force
law is softened with the proper softening length about 0.004 the final
virial radius. Each simulation has evolved 10,000 time steps by the
leap-frog algorithm.

\section{Density Profiles}

\begin{figure}[t]
\begin{center}
\epsfile{file=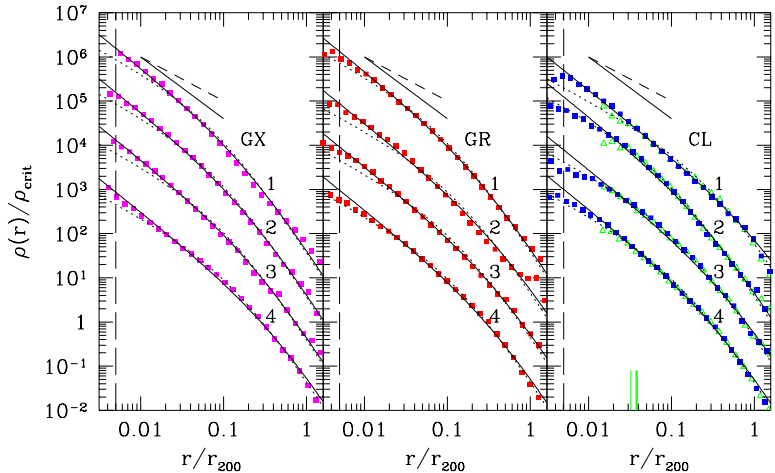,width=0.70\textwidth}
\end{center}
\caption{The density profiles of the simulated halos of galaxy ({\it
left}), group ({\it middle}), and cluster ({\it right}) masses.  The
solid and dotted curves represent fits of $\beta=1.5$ and $\beta=1$
respectively.  The vertical dashed lines indicate the force softening
length. The open triangles in the right panel show the results of the
cosmological simulation with the force softening marked by the long 
ticks at the bottom. The density value of each halo is multiplied by 1,
$10^{-1}$, $10^{-2}$, $10^{-3}$ from top to bottom in each panel.}
\label{fig:haloprof}
\end{figure}

\begin{figure}[t]
\begin{center}
\epsfile{file=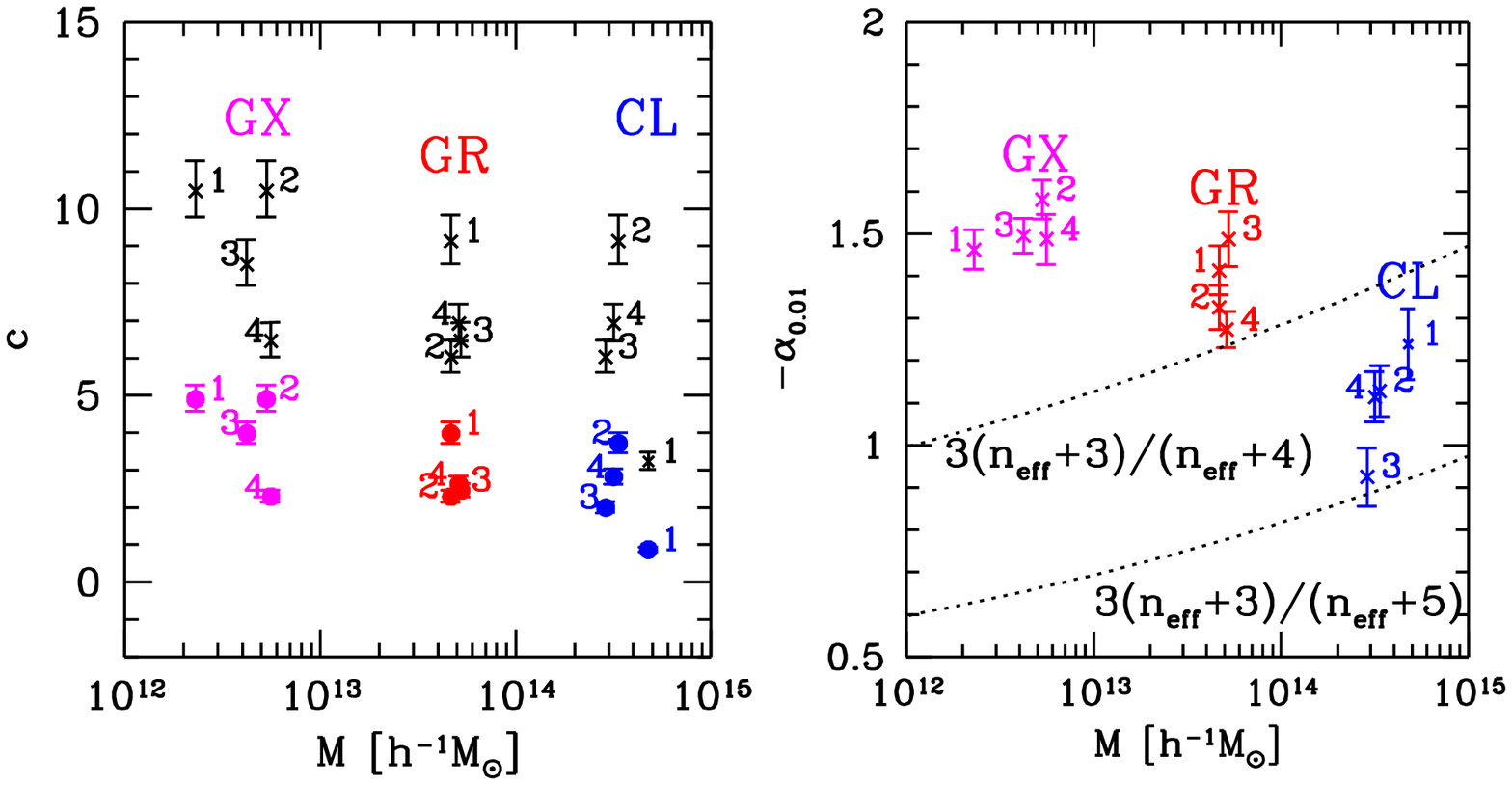,width=0.76\textwidth}
\end{center}
\caption{{\it Left panel:} the concentration parameters for each halo
for $\beta=1.5$ (filled circles) and for $\beta=1$ (crosses).  Numbers
labeling each symbol correspond to the halo ID. {\it Right panel:}
Power-law index of the inner region ($0.007 <r/r_{200}<0.02$) as a
function of the halo mass. The upper and lower dotted curves bound the
current analytical predictions (see Jing \& Suto 2000 for a
discussion)}
\label{fig:haloslope}
\end{figure}

As an important application, we have examined the radial density
profiles (Jing \& Suto 2000) which are plotted in Figure
\ref{fig:haloprof}.  The inner slope of the profiles, however, is
generally steeper than the NFW value, $-1$, in agreement with the
previous findings of Fukushige \& Makino (1997) and Moore et
al. (1998).  We have fitted the profiles to $\rho(r)\propto
r^{-\beta}(r+r_s)^{-3+\beta}$ with $\beta=1.5$ (the solid lines) and
$\beta=1$ (NFW form; the dotted curves) for $0.01r_{200}\le r \le
r_{200}$, where $r_{200}$ is the radius within which the spherical
overdensity is $200 \rho_{\rm crit}(z=0)$.  The resulting
concentration parameter $c$, defined as the $r_{200}/r_s$, is plotted
in the left panel of Figure \ref{fig:haloslope}. This is the most
accurate determination of the concentration parameter for the LCDM
model. There exists a significant scatter among $c$ for similar mass
(Jing, 1999), and a clear systematic dependence on halo mass (NFW,
Moore et al. 1999).

Our most important result is that the density profiles of the 4
galactic halos are all well fitted by $\beta=1.5$, but those of the
cluster halos are better fitted to the NFW form $\beta=1$. This is in
contrast with Moore et al.(1999) who concluded that both galactic and
cluster halos have the inner density profile $\rho(r)\propto
r^{-1.5}$, despite that they considered one cluster-mass halo
alone. In fact, our current samples can address this question in a
more statistical manner. CL1 has significant substructures, and the
other three are nearly in equilibrium. Interestingly the density
profiles of CL2 and CL3 are better fitted to the NFW form, and that of
CL4 is in between the two forms. The density profiles of the group
halos are in between the galactic and cluster halos, as expected. One
is better fitted to the NFW form, whereas the other three follow the
$\beta=1.5$ form.

\section{Conclusion and Discussion}

We have presented the first study which simulates fifteen dark
halos with about a million particles.  This enables us to address the
profile of the halos with unprecedented accuracy and statistical
reliability.  While qualitative aspects of our results are not
inconsistent with those reported by Moore et al. (1999), our larger
sample of halos provides convincing evidence that {\it the form} of
the density profiles is not universal; instead it depends on halo
mass. Since mass and formation epoch are linked in hierarchical
models, the mass dependence may reflect an underlying link to the age
of the halo. Older galactic halos more closely follow the $\beta=1.5$
form while younger cluster halos have shallower inner density profiles
fitted better by the NFW form.

\vspace{1pc}\noindent
Y.P.J. gratefully acknowledges a JSPS fellowship.  Numerical
computations were carried out on VPP300/16R and VX/4R at ADAC, NAOJ.

\section{References}

\vspace{1pc}
\re
  Fukushige, T., \& Makino, J. 1997, ApJ, 477, L9
\re
Kravtsov, A. V., et al. 1998, ApJ, 502, 48
\re
  Jing, Y.P. \& Suto, Y. 1998, ApJ, 494, L5
\re
  Jing, Y.P. \& Suto, Y. 2000, ApJ, 529, L69
\re
  Jing, Y. P. 2000, ApJ, in press (astro-ph/9901340)
\re
  Moore, B. et al. 1998, ApJ, 499, L5
\re
  Moore, B., et al. 1999, MNRAS, submitted, astro-ph/9903164
\re
  Navarro, J.F., Frenk, C.S., \& White, S.D.M. 1997, ApJ, 490, 493

\end{document}